\documentclass[graybox]{svmult}

\usepackage{mathptmx}       
\usepackage{helvet}         
\usepackage{courier}        
\usepackage{type1cm}        
\usepackage{makeidx}         
\usepackage{graphicx}        
\usepackage{multicol}        
\usepackage[bottom]{footmisc}

\usepackage{natbib}
\makeindex             

\def\mnras{MNRAS}
\def\aap{A\&A}
\def\apj{ApJ}
\def\apjl{ApJ}
\def\aj{AJ}

\def\pasp{PASP}
\def\apjs{ApJS}
\def\aaps{A\&AS}
\def\kms{{\rm\,km\,s^{-1}}}

\begin{document}

\title*{The outer halos of elliptical galaxies}
\author{Ortwin~Gerhard}
\institute{MPE, Giessenbachstrasse, D-85748 Garching, Germany
           \email{gerhard@mpe.mpg.de}
}
%
%
\maketitle

\abstract*{Recent progress is summarized on the determination of the
  density distributions of stars and dark matter, stellar kinematics,
  and stellar population properties in the extended, low surface
  brightness halo regions of elliptical galaxies.  With integral field
  absorption spectroscopy and with planetary nebulae as tracers,
  velocity dispersion and rotation profiles have been followed to
  $\sim 4$ and $\sim 5-8$ effective radii, respectively, and in M87
  to the outer edge at $\sim 150$ kpc. The results are generally
  consistent with the known dichotomy of elliptical galaxy types, but
  some galaxies show more complex rotation profiles in their halos and
  there is a higher incidence of misalignments, indicating
  triaxiality. Dynamical models have shown a range of slopes for the
  total mass profiles, and that the inner dark matter densities in
  ellipticals are higher than in spiral galaxies, indicating earlier
  assembly redshifts. Analysis of the hot X-ray emitting gas in X-ray
  bright ellipticals and comparison with dynamical mass determinations
  indicates that non-thermal components to the pressure may be
  important in the inner $\sim10$ kpc, and that the properties of
  these systems are closely related to their group environments. First
  results on the outer halo stellar population properties do not yet
  give a clear picture.  In the halo of one bright galaxy, lower
  [$\alpha$/Fe] abundances indicate longer star formation histories
  pointing towards late accretion of the halo. This is consistent with
  independent evidence for on-going accretion, and suggests a
  connection to the observed size evolution of elliptical galaxies
  with redshift.}

\abstract{Recent progress is summarized on the determination of the
  density distributions of stars and dark matter, stellar kinematics,
  and stellar population properties in the extended, low surface
  brightness halo regions of elliptical galaxies.  With integral field
  absorption spectroscopy and with planetary nebulae as tracers,
  velocity dispersion and rotation profiles have been followed to
  $\sim 4$ and $\sim 5-8$ effective radii, respectively, and in M87
  to the outer edge at $\sim 150$ kpc. The results are generally
  consistent with the known dichotomy of elliptical galaxy types, but
  some galaxies show more complex rotation profiles in their halos and
  there is a higher incidence of misalignments, indicating
  triaxiality. Dynamical models have shown a range of slopes for the
  total mass profiles, and that the inner dark matter densities in
  ellipticals are higher than in spiral galaxies, indicating earlier
  assembly redshifts. Analysis of the hot X-ray emitting gas in X-ray
  bright ellipticals and comparison with dynamical mass determinations
  indicates that non-thermal components to the pressure may be
  important in the inner $\sim10$ kpc, and that the properties of
  these systems are closely related to their group environments. First
  results on the outer halo stellar population properties do not yet
  give a clear picture.  In the halo of one bright galaxy, lower
  [$\alpha$/Fe] abundances indicate longer star formation histories
  pointing towards late accretion of the halo. This is consistent with
  independent evidence for on-going accretion, and suggests a
  connection to the observed size evolution of elliptical galaxies
  with redshift.}

\section{Introduction}
\label{sec:1}

There are indications that the outer halos of bright elliptical
galaxies formed later and through different processes than their inner
parts. It is clearly important to establish whether this is true for
some, most, or all ellipticals. Signatures of the formation processes
may be preserved longer in the outer halos than in the well-mixed
inner parts, because of the longer dynamical time-scales at large
radii. Owing to the low surface brightness in these halos, the outer
kinematics and stellar population properties can be best studied
in local galaxies. Combining the results with predictions from
simulations and with high-redshift studies of ellipticals one may
expect to obtain a global picture of halo formation in elliptical
galaxies.  The following sections will discuss some recent results on
the stellar distribution and kinematics in the outer halos (\S2), the
dark matter distribution and dynamics (\S3), and finally the stellar
populations and assembly history of nearby elliptical galaxy halos
together with related results from high-redshift studies and
simulations (\S4).

The subject of the outer halos of ellipticals is one that has
benefitted enormously from the vision of Ken Freeman, whose scientific
work we celebrate with this conference. Ken was one of the first to
realize the potential of using planetary nebulae (PNe) as kinematic
tracers in regions where the surface brightness is too low for
traditional absorption line spectroscopy. In 1987, in the course of a
project with X.~Hui, H.~Ford and M.~Dopita, he measured the first PN
radial velocities in the halo of Centaurus A, the nearest elliptical
galaxy, using a multi-fiber instrument at the Anglo-Australian
Telescope. Together with later AAT and CTIO measurements, they were
able to obtain 433 PN velocities in this galaxy \citep{Hui+95}.
Together with M.~Arnaboldi and others, he extended this work to the
fainter PNe in the halos of more distant ellipticals such as NGC 1399,
M86, and NGC 1316 \citep{Arnaboldi+94,Arnaboldi+96,Arnaboldi+98}, in
order to constrain the kinematics and dark matter mass distribution in
these galaxies.

The potential shown by these papers led to the idea of a
special-purpose instrument, the Planetary Nebula Spectrograph (PNS),
which is based on the principle of using counterdispersed imaging to
simultaneously find PNe and measure their radial velocities
\citep{Douglas+97,Douglas+02}. This instrument is currently mounted at
the William Herschel Telescope in La Palma. As two highlights, some
2600 PN velocities were measured in nearby M31 \citep{Merrett+06}, and
within a still ongoing project $\sim$100-200 PN velocities have so far
been obtained for each of a sample of some 15 ellipticals
\citep{Coccato+09}.  The 1996 paper on M86 also reported three
discrepant PN velocities, ascribed to the intracluster stellar
population in Virgo. Since then, Ken has enthousiastically
participated in using PNe to study the dynamics of intracluster stars
in nearby galaxy clusters \citep{Freeman+00}; see also M.~Arnaboldi's paper in this volume.
For the first multi-slit imaging observations of PNe in the Coma
cluster \citep{gerhard+05} he went observing at the Subaru telescope.
Ken is a coauthor of some 30 refereed papers related to the use of PNe
for kinematic measurements, including also the Galactic bulge
\citep{Beaulieu+00} and, again, Centaurus A \citep{Peng+04}.

\section{Density distribution and kinematics of outer halo stars}
\label{sec:2}

Accurate brightness profiles over large radius ranges were constructed
by \citet{Kormendy+09} for Virgo cluster elliptical and spheroidal
galaxies. The brightest elliptical galaxies in this sample have
profiles characterized by large S\'ersic indices $n\simeq 4-12$, large
effective radii $R_e>10$ kpc at faint surface brightnesses,
\hbox{$\mu_{e,V}\simeq 23-25$}, and central cores (depressions relative to the
S\'ersic profile). Fainter ellipticals have S\'ersic indices $n\sim$2-4,
$R_e$'s of a less than a few kpc, $\mu_{e,V}\simeq 19-22$, and central
cusps (extra light relative to the S\'ersic profile). While the
dichotomy in the central profiles of both classes is clear and is
their primary distinction, there are some transition cases in the
large-radius profiles.  The important point for this talk is that the
(bright) core ellipticals have the most extended luminous halos, and
that within each of the two classes $n$ (extent) does not correlate
with luminosity \citep{Caon+93,Kormendy+09}.

Outer halo kinematics based on PN velocities were analyzed by
\citet{Coccato+09} for a sample of 15 early-type galaxies. These data
typically reach to $\sim 5R_e$ and in some galaxies to $8R_e$. The
derived profiles of PN number distribution and kinematics agree with
surface brightness and absorption line kinematics where the data sets
overlap, within their respective errors. The radial profiles of mean
rms velocity $v_{\rm rms}$ for this sample fall within two groups,
with part of the galaxies having slowly decreasing $v_{\rm rms}(R)$
and the remainder having steeply falling $v_{\rm rms}(R)$. There is a
large overlap of these groups with the core and cusp ellipticals,
respectively \citep{Kormendy+09,Hopkins+09a, Hopkins+09b}. The outer
halos of ellipticals tend to be more rotationally dominated than the
central parts. The slow-rotator versus fast-rotator classification of
\citet{Emsellem+07} based on inner Sauron data is largely preserved in
the halos; however, some more complex profiles of specific angular
momentum parameter $\lambda_R$ are seen in the halos. Slow rotators
typically reach $\lambda_R\simeq 0.1$ in their outer parts. Lastly,
twists and misalignments in the velocity fields, commonly interpreted
as signs of triaxiality, seem to be more frequent at large radii
\citep{Coccato+09}.

\citet{Proctor+09} have pioneered a technique to measure absorption
line kinematics for the integrated halo stellar population using data
from multi-object slitlet spectroscopy, and to construct 2D velocity
fields to $3R_e$ from these data.  They find a variety of stellar
rotation profiles beyond $1R_e$, constant, decreasing, and increasing,
confirming that the fast-slow rotator classification may be more
complicated in the halos than within $R_e$. Halo kinematics obtained
with integral field units (IFUs) such as Sauron and VIRUS-P reach
somewhat further, to $\sim 4R_e$ \citep{Murphy+07,Weijmans+09}.  In
these studies, the signal from many lenslets or fibers is co-added to
increase the signal-to-noise.  Because of the limited field-of-view,
several IFU fields may need to be exposed.  For the galaxies where
kinematics is available both from these absorption line techniques and
from PNe, the velocity dispersion profiles agree within the errors. The
integrated light techniques yield measurements also of the line
profile shape parameters, giving valuable constraints on the
anisotropy of the halo orbit distribution.  These are harder to obtain
from PNe unless large samples can be obtained. Since PNe can be found
out to larger radii, the best dynamical constraints will come from
combining both techniques.

The extreme outer halo of the Virgo-centric giant elliptical galaxy
M87 was studied by \citet{Doherty+09}. They used small samples of PNe
in fields at 60 and 150 kpc to measure the velocity dispersion, after
removing Virgo intracluster stars. PNe centered around the systemic
velocity of M87 were found only out to 150 kpc, but not further, and
the velocity dispersion profile was found to decline steeply towards
the outer edge.  Given that the X-ray-determined mass profile is
rising steeply at these radii and is dominated by the cluster, this is
dynamically possible only if the halo of M87 is truncated.
\citet{McNeil+10} studied the central bright galaxy in the Fornax
cluster, NGC 1399. Again using PNe, they found that a component of
low-velocity PNe (by $\sim 800 \kms$) is superposed on the main galaxy
population. When this component is removed, the velocity dispersion
profile of NGC 1399 is flat at $\sim 200\kms$ from PNe and red
globular clusters out to 80 kpc. Both studies show that the outermost
halos of elliptical galaxies are in interaction with the environment,
and that discrete tracers are important to disentangle the different
(mixed and unmixed) components in the velocity distribution.

\section{Dark matter and dynamics}
\label{sec:3}

The dark matter distribution in elliptical galaxies can be studied
with a variety of techniques, including strong and weak lensing, X-ray
emitting hot gas, and dynamics. The lensing results are discussed in
K.~Kuijken's talk in this volume; in brief, while with weak lensing data the
dark halos of elliptical galaxies can be followed to hundreds of kpc,
the strong lensing analyses give accurate mass measurements within the
Einstein radius, and together with central velocity dispersions 
constrain the slopes of the mass density profiles to an
average value of very nearly $-2$ (isothermal), with small scatter.
 
Early dynamical analysis of stellar velocity dispersions and line
profile shape data with non-parametric spherical models first
indicated that the inner circular velocity curves (CVC) of elliptical
galaxies are flat to $\sim 10\%$, out to approximately 1-2 $R_e$
\citep{Kronawitter+00, Gerhard+01}. Recent work by \citet{Thomas+07}
on a sample of ellipticals in the Coma cluster with slightly more
extended data (1-3 $R_e$) has shown more varied CVC slopes, with some
falling, some flat, and some rising. Using axisymmetric Schwarzschild
models to constrain the dark matter halos in the Coma ellipticals,
\citet{Thomas+09} found that the dark matter densities in these
galaxies are on average $7\times$ higher than in spiral galaxies of
the same luminosity, and $13\times$ higher than in spirals of the same
baryonic mass, consistent with the earlier analysis of round nearby
galaxies by \citet{Gerhard+01}. Baryonic contraction is not sufficient
to explain the difference, implying that the inner halos of
ellipticals presumably formed earlier than those of spiral galaxies.
If one assumes that the halo densities measure the density of the
Universe at the time of collapse \citep{Gerhard+01}, the assembly
redshift $(1+z_{\rm ass})$ for the Coma ellipticals is twice that for
typical spiral galaxies, and puts their $z_{\rm ass}=1...3$
\citep{Thomas+09}, with brighter ellipticals assembling later.

Several intermediate-luminosity ellipticals have steeply falling outer
PN velocity dispersion profiles \citep{Romanowsky+03,Coccato+09}.  The
prototype galaxy for this class is NGC 3379, for which some 200 PN
velocities are available. It was thus argued 
\citep{Romanowsky+03,Douglas+07} that this galaxy might be `naked',
i.e., lack a significant dark matter halo.  Extensive recent
dynamical modeling by \citet{deLorenzi+09}, using the made-to-measure
particle code NMAGIC \citep{deLorenzi+07}, has shown that a variety of
dark matter halos are allowed by the data, implying radially anisotropic
orbit distributions, and that the no-dark matter
halo is ruled out \citep[see also][]{Weijmans+09}. However, the
allowed halo density distributions imply that the CVC in this galaxy
falls by at least 25\% between the center and $2R_e$, i.e.\ they are
of significantly lower density than would be required for a flat CVC.
Similar results were found for two other galaxies of this class,
NGC 4697 and NGC 4494 \citep{deLorenzi+08,Napolitano+09}.

There is a long history of mass determination in X-ray bright
elliptical galaxies, using the hot gas as a tracer in assumed
hydrostatic equilibrium in the gravitational potential
\citep[e.g.][]{Nulsen+95,Humphrey+06,Fukazawa+06,Nagino+09}. Based on
a comparison of the gravitational potentials and circular velocities
inferred from X-rays and from dynamics, \citet{Churazov+08,
  Churazov+10} argued that the non-thermal sources of pressure
contribute on average 20-30\% of the thermal gas pressure.
\citet{Das+10} recently devised a non-parametric Bayesian technique to
determine CVCs and confidence ranges from deprojected hot gas pressure
and temperature profiles. Comparing with dynamical mass
determinations, they also find differences indicating that non-thermal
pressure sources are important in the central 10 kpc of their analyzed
six galaxies. Other possible interpretations for these differences are
violations of hydrostatic equilibrium or biases introduced by the
assumptions in the dynamical models; this is an important issue that
needs to be clarified.  The dark matter mass fractions inferred by
\citet{Das+10} are $35-80\%$ at $2R_e$, rising to $80-90\%$ at the
outermost radii.  \citet{Das+10} also find that for all of their
galaxies the outer CVCs are rising, implying steeper than isothermal
mass profiles. Furthermore, the inferred circular velocities and the
luminosities of their galaxies are found to correlate with the
velocity dispersion of the environment.  Both results suggest that the
properties of X-ray bright ellipticals are closely related to their
group environments.

\section{Stellar population and assembly history}
\label{sec:4}

Studies of the stellar populations in the outer halos of early-type
galaxies may give important constraints on their assembly history. Due
to the faint surface brightnesses in the outer regions this work is
only in its beginning, requiring large telescopes and/or special
techniques.

In a recent study, \citet{Coccato+10a} used deep medium resolution
spectroscopy with FOCAS at the Subaru telescope to measure Lick
line-strength indices far into the halo of one of the two Coma
brightest cluster galaxies (BCGs), NGC 4889. Combining these with
literature data, \citet{Coccato+10b} constructed radial profiles of
metallicity, [$\alpha$/Fe] abundance ratio and age, from the center
out to $\sim 60$ kpc ($\sim4 R_e$). These profiles show evidence for
different chemical and star formation histories for stars inside and
outside $\sim 1.2 R_e = 18$ kpc radius. The inner regions have a steep
metallicity gradient and high [$\alpha$/Fe] at $\sim2.5$ solar value,
pointing to a rapid formation process. This is consistent with a
quasi-monolithic, dissipative merger collapse at early redshifts
\citep{Kobayashi04}, followed by one or at most a few dry mergers
between several such units so that the original steep metallicity
gradient \citep{Chiosi+02,Pipino+08} was only partially erased
\citep{diMatteo+09}. In the halo of NGC 4889, between $18$ and $60$
kpc, the metallicity is near-solar with a shallow gradient, while
[$\alpha$/Fe] shows a strong negative gradient, reaching solar values
at 60 kpc, and the inferred ages are 9-13 Gyr. This argues for later
accretion of stars from old systems with more extended star formation
histories \citep{Matteucci94}, presumably through minor mergers from
the galaxy's group environment \citep{Abadi+06,Naab+09}.

\citet{Rudick+10} study the colour distribution in the outer halo of
M87 and the adjacent intracluster light. They find that the M87 halo
becomes bluer with radius to the outermost radii, and argue that the
common colours of the outer halo and of some surrounding intracluster
tidal features suggest that the galaxy's envelope may have formed from
similar streams.  \citet{Foster+09} used the technique from
\citet{Proctor+09} to study three galaxies. Their most extended
dataset is for NGC 2768 which they suggest, based on comparison with
simulations by \citet{Hopkins+09a}, is consistent with a dissipative
merger origin. The two other datasets have too large errors beyond
$R_e$ to allow firm conclusions.  \citet{Weijmans+09} find from
co-added Sauron field data that the metallicity gradients in the two
intermediate luminosity ellipticals NGC 3379 and NGC 821 remain
approximately constant to $4R_e$, with old and subsolar metallicity
populations in the halos. Thus outer halo stellar population studies
to date do not yet combine to a clear picture.

The bimodal stellar populations found by \citet{Coccato+10b} in the
core and outer halo of NGC 4889 are consistent with recent results on
the size evolution of early-type galaxies (ETGs) with redshift
\citep[e.g][]{Daddi+05,Trujillo+07,Cimatti+08}, such that ETGs at
$z\sim1$ ($z\sim2$) have sizes a factor of 2 ($3-5$) smaller than ETGs
with similar mass today
\citep{vanderWel+08,vanDokkum+08,Saracco+09}. Recently,
\citet{vanDokkum+10} find from stacked rest-frame R-band images that
massive ETGs have nearly constant mass inside 5 kpc with redshift, but
increase their envelope mass by a factor $\sim 4$ since $z=2$, with
effective radius evolving as $R_e\propto (1+z)^{-1.3}$. Minor mergers
may play an important role in this process \citep{Naab+09}, and the
accreted galaxies are likely to have gone through longer star
formation histories, leading to less $\alpha$-enriched stellar
populations as was found in the halo of NGC 4889.  Further work on the
stellar populations in the outer halos of local elliptical galaxies
may become a valuable tool to understand what part of the
elliptical galaxy population participates in the observed size
evolution.

{\bf Acknowledgments:} First, it is a pleasure on this occasion to thank
Ken Freeman for the long-standing collaboration and the many
stimulating discussions over the years. I would also like to thank
David Block and his team for organizing this memorable conference in
Ken's honour, and for all the extra work they did in order to be able
to hold it in this fantastic location in the Namibian desert.

\end{document}